*Reference:*
*van Haren, H. (2023). Direct observations of general geothermal convection in deep Mediterranean waters. Ocean Dynamics, 73, 807-825.*# Direct observations of general geothermal convection in deep Mediterranean waters

**Hans van Haren**

NIOZ Royal Netherlands Institute for Sea Research, P.O. Box 59, 1790 AB Den Burg, the Netherlands.
hans.van.haren@nioz.nl1

**Abstract**

Like elsewhere in the deep-sea, life in the deep Mediterranean depends on turbulent exchange across the stable vertical density stratification for supply of nutrients and oxygen. Commonly modelled, turbulent exchange is inversely proportional to the stratification rate. However, this proportionality depends on the particular turbulence type, whether it is driven by vertical current differences (shear) or by buoyancy (convection). While shear-turbulence is well observed in stratified seas, direct observations of convection-turbulence are limited. In this paper, high-resolution moored temperature observations show that Mediterranean Sea waters are not stagnant in the lower 109 m above the seafloor at 2480 m, although variations are in the range of only 0.0001-0.001 °C. In winter, convection-turbulence is regularly observed. Fortnightly averaged spectra show a collapse to the inertial-subrange scaling of dominant shear-turbulence for data from about 100 m above the seafloor, and to the buoyancy-subrange scaling of dominant convection-turbulence at about 10 m above the seafloor. Time-depth images reveal details of convection-turbulence driven from below, which is considered primarily due to general geothermal heating through the Earth crust not related to volcanic vents. When its observation is not masked by (sub-)mesoscale eddies that advect warmer waters from above, the geothermal heat flux matches the deep-sea turbulence dissipation rate, if in the calculations a mixing efficiency of 0.5 is taken typical for natural convection, integration is over 250 m above the seafloor as confirmed from shipborne CTD, and if maximum 2-m-scale buoyancy frequency replaces its 100-m-scale mean equivalent.

**Keywords** Deep Western Mediterranean; High-resolution moored temperature observations; Geothermal heating outside volcanic vents; Convection turbulence directly observed

# 1 Introduction

In geophysical fluid dynamics environments, the Reynolds number Re = UL/ν, where U denotes velocity, L a length scale and ν = $10^{-6}$ $m^2$ $s^{-1}$ kinematic viscosity, is generally high, with typical values between $10^6$ < Re < $10^7$. This is because of the large spatial scales, so that the flow is turbulent nearly always and nearly everywhere. The flow-turbulence is important for the exchange and redistribution of matter, for example of nutrients and suspended sediment, as this mechanical irreversible process acts orders of magnitude faster and over larger scales than molecular diffusion.

Of the geophysical environments, the ocean-interior is particular because it demonstrates mainly 'stratified turbulence': Virtually everywhere it is stably stratified in density upon which turbulence acts as a destabilizer. The primary agent determining density variations, and thus the omnipresence of stratification, is temperature and the ocean is heated (by the sun) and cooled from the same geopotential at its top (Sandström 1908). It forms the major contrast in dynamics of oceanography and meteorology, as the atmosphere is heated from below and cooled at its top (Munk and Wunsch 1998). At greater depths in the ocean, stratification becomes weaker but waters are still turbulent.

The vertically density-stratified environment supports two main turbulence types: shear-turbulence that is induced via differential flow causing friction with roll-up of isopycnals as in Kelvin-Helmholtz instabilities KHi, and convection-turbulence that is induced via unstable buoyancy gradients as in Rayleigh-Taylor instabilities RTi. In general, both types intermingle as primary roll-up generates local secondary RTi and convection-plumes generate secondary KHi on their sides. In the mean, the mechanical overturning of stably stratified waters dominated by primary shear-turbulence has a mean mixing efficiency of about 0.2 (e.g., Oakey 1982; Mashayek et al. 2021), although the spread of values is large and although efficiency reduces to zero in fully mixed 'homogeneous' layers. Convection-turbulence, 'naturally' buoyancy-driven by gravitationally unstable conditions of less dense (warmer, fresher) waters underneath denser (colder, saltier) waters, has a high mixing efficiency of 0.5 (Dalziel et al. 2008; Gayen at al. 2013;

Ng et al. 2016). Convection is considered to be turbulent when it is considerably larger than conduction (e.g., White 1984), when the Nusselt number $Nu = Q/\Delta T/(k/L) > 100$, where Q denotes the heat flux, $\Delta T$ the temperature difference and $k = 0.6$ W/m·°C the thermal conductivity. Equivalently above a flat boundary, Rayleigh convection becomes turbulent when the Rayleigh number $Ra = (g\beta/\nu\alpha)\Delta T L^3 > 10^7$ (Foster 1971), where g denotes the acceleration of gravity, $\beta = 2\times10^{-4}$ °C$^{-1}$ the thermal expansion coefficient and $\alpha = 1.4\times10^{-7}$ m$^2$ s$^{-1}$ the thermal diffusivity. Shear-turbulence is considered the dominant agent in the well stratified ocean-interior. It is mainly set-up via internal wave breaking (Eriksen 1982; Thorpe 2010), while convection-turbulence is considered dominant in the atmosphere and only minor in the ocean.

Although convection-turbulence is not often dominant in the sea, exceptions are the near-surface 10-50-m deep layer during nighttime cooling (e.g., Brainerd and Gregg 1995), and, much more localized both in space and time, dense water formation in polar and Mediterranean Sea areas during particular cooling and evaporation events in late winter (e.g., Marshall and Schott 1999). A more continuous turbulent convection is expected from the heating of deep-sea waters from below, by geothermal leakage from the Earth mantle via conduction through the Earth crust (e.g., Davies and Davies 2010; Wunsch 2015). General seafloor heating is meant here, not heating via highly localized 1-m diameter small volcanic vents (e.g., Wenzhöfer et al. 2000). Whilst heat-flow measurements are a common observable for geophysicists, only indirect oceanographic observational evidence of general geothermal heating has been found so far (e.g., Emile-Geay and Madec 2009). Multiple profiling has revealed evidence of geothermal convective heating in the Black Sea (Murray et al. 1991) in a layer of up to 450 m above the seafloor (Kelley et al. 2003), and which dominates over the bottom heat inflow through the narrow Bosporus Strait (Stanev et al. 2021).

To put various sources of heat and energy in perspective, the sea-surface receives about 100,000 TW (1 TW = $10^{12}$ W), i.e. 300 W m$^{-2}$, of short-wave radiation by the sun during daytime, mankind uses nowadays about 17 TW of energy, omnipresent sea-interior internal wave kinetic energy amounts about 3 TW, and quasi-permanent general geothermal heating of deep-sea waters totals about 35 TW, or 0.1 W m$^{-2}$ (Wunsch 2015).

Thus far, no direct observations of convection-turbulence have been made in winter cascades of vertical (convection-)'plumes' of up- and down-going waters during deep dense water formation (Thorpe 2005). Even diurnal ocean convection lacks direct observations of vigorous convection-plumes, despite sub-hourly turbulence measurement efforts (e.g., Lombardo and Gregg 1989; Brainerd and Gregg 1995). High-resolution sampling appears difficult near the sea-surface.

However, also turbulent convection due to geothermal heating has not been directly observed in the relatively quiescent waters above a deep seafloor, despite the weak, compared to outward nighttime long-wave radiation near the sea-surface, but non-negligible heat-input from below. The reasons are obvious, as such direct observations require instrumentation that withstand high deep-sea pressures and that collect data over a wide range of turbulent overturning scales. Thereby, (at least) 1-m spatial scales are to be resolved over a range O(100) m, and 1-s time scales over at least the buoyancy and preferably inertial periods of variation. Data are to be collected in thus weakly stratified waters that both periods of variation are O(10) h and temperature variations are <10$^{-4}$ °C over a range of <10$^{-3}$ °C. One needs sensitive and numerous instrumentation.

In the ocean, near-homogeneous conditions with weak stratification in which buoyancy frequency N ≈ f, the local inertial frequency of Earth rotation or Coriolis parameter, implies a vertical density variation of only 0.0001 kg m$^{-3}$ over a range of 100 m of water. It thus poses a challenge on every instrument that is moored in such waters that, for example, occur in the deep-sea, for studying turbulence details. The high-resolution temperature (T)-sensors used here are no exception, and although their noise levels are smaller than 0.0001 °C their data require elaborate post-processing. Data from such moored T-sensors provide different spectra in turbulence transition sub-ranges of buoyancy, for active scalar (anisotropic stratified convection-turbulence;

Bolgiano 1959; Zongo and Schmitt 2011; Pawar and Arakeri 2016), and of universal equilibrium inertial, for passive scalar (isotropic shear-turbulence; Ozmidov, 1965; Tennekes and Lumley 1972; Warhaft 2000).

The small temperature (density) variations are expected, because in more strongly stratified waters geothermal heating would be too weak to overcome stability, like near-surface convection is generally blocked during daytime heating from above. Geothermal heating is thus not expected to be observable over sloping ocean topography where internal wave breaking dominates the turbulent mixing (e.g., Eriksen 1982) that is followed by rapid restratification due to the sloshing waves (e.g., Winters 2015).

In this paper, general geothermal heating is presented in 2500-m deep Western Mediterranean waters over a flat seafloor as directly observed using an array of high-resolution moored T-sensors. Previous attempts using the same sensors in weakly stratified waters above a 4000-m deep East-Pacific flat seafloor (van Haren 2020) and in deep alpine-lake Garda (van Haren and Dijkstra 2021) did not show clear evidence of convection-turbulence induced by geothermal heating. These weakly stratified waters were dominated by convection-turbulence that was induced by downward acceleration of internal waves supported by stratified waters above. Other processes that, like stratification, (sub-)mesoscale eddies, and internal waves, dominate over geothermal heating, and thus mask its observation in temperature measurements, are expected to regularly occur in the deep North-Western Mediterranean.

The near-coastal NW-Mediterranean is known for a steep continental slope, strong boundary flows of which the intensity varies seasonally with large mesoscale and sub-mesoscale eddy-activity, occasional deepwater formation during some late-winters and large near-inertial internal waves, even though tidal motions are weak (Crépon et al. 1989; Albérola et al. 1995; Millot 1999; Testor and Gascard 2006). In these $N=f$ waters, internal waves have the property of large vertical columns like in eddies (Straneo et al. 2002).

In the observational area, the average amount of the vertical flux attributed to geothermal heating is 100±30 mW m$^{-2}$ (Pasquale et al. 1996), which is close to the average value yielding a global seafloor heating of 35 TW as the ocean surface amounts 3.6×10$^{14}$ m$^2$. This geothermal heat-flux value is about one-third of that over the Mid-Atlantic Ridge and about triple that of the Eastern-Mediterranean. Bethoux and Tailliez (1994) calculated a yearly temperature increase of 0.0068°C over 100 m above the Western Mediterranean seafloor due to geothermal heating. For the central Western Mediterranean, shear-induced turbulent mixing was found to be about half the size of geothermal convection-turbulence, using extensive shipborne microstructure profiling (Ferron et al. 2017). Hereby, geothermal heating was considered to reach up to 1200 m from the seafloor. This 1200-m height seems too high for the present mooring area, where the proximity to the continental slope is expected to provide regular and sufficient stratification by guiding the boundary flow and associated eddies for masking convection, from spring-autumn (van Haren 2023).

## 2 Technical details

To evaluate various turbulence processes in weakly stratified deep-sea, a small-scale three-dimensional (3D) mooring array of high-resolution temperature sensors is used to investigate the contributions of anisotropic, vertically flattened, stratified turbulence and of isotropic turbulence to convection-turbulence. Spectral information is mainly used to distinguish these types of turbulence. Shipborne observations may provide higher vertical resolution but are inadequate to resolve temporal variations. Such observations are used to provide additional information on temperature and on other parameters like salinity and density.

### 2.1 3D mooring array of temperature sensors

Few 3D instrumented devices exist to study internal waves and turbulence (Thorpe, 2010). He proposed a freely floating device with 3 lines 10 m apart and 20 m tall holding 200-300 temperature sensors. In practice however, such a device causes insurmountable problems to have it float at 600-1000 m below the sea-surface as suggested. To prevent distortion, sufficient tension of about 1 kN per line should be applied, which can only be achieved via sufficient weight below and buoyancy above. The hypothetical heavy freely floating device should be controlled to a net neutral buoyancy, a difficult trimming of 0.01 kN over a mean of several kN. Instead, a multiple-line mooring is used here (Fig. 1a, b).

The NIOZ (Royal Netherlands Institute for Sea Research) five-line (5-L) mooring is, when fold-up, a 6-m tall and 3-m diameter high-grade aluminum structure that can contain 520 'NIOZ4' high-resolution T-sensors (van Haren et al. 2016). Completely stretched, it measures about 110 m in height and 5.6x5.6 m horizontally and the five mooring lines are 4, 5.6 and 8 m apart thereby spanning a small 3D deep-sea volume of 3500 $m^3$. 5-L consists of two support frames of 1.7x1.7 m each holding a set of four arms 3.3 m long. Four instrumented cables connect the corner tips of the upper and lower sets of arms ('line-1...4'); a central instrumented cable connects the upper and lower inner frames ('line-c'). Each line is held under 1 kN of tension by heavy top-buoyancy when on the seafloor, and by a 9-kN anchor-weight during overboard operations and free-fall deployment. The tension was sufficient to keep mooring-rotations to within ±3° under drag of 0.35 m $s^{-1}$ water-flow speeds.

For the Western Mediterranean deployment, a total available amount of 340 T-sensors is taped to 5-L: 104 sensors at 1.0 m intervals to line-c with the lowest sensor at 5 m above the seafloor, 53 sensors at 2.0 m intervals to each of line-1…4, 4 sensors at 1.0 m intervals to each corner-bottom-weight-line. As a result, 52 overlapping vertical positions exist between 5 and 108 m above the seafloor at which all five lines have a sensor. The lowest sensors were mounted near the top of each corner-weight (Fig. 1b) at about 0.5 m above the seafloor. The somewhat bulky 0.1-m diameter 0.25-kN weighing lead-filled corner-weight pipes stuck several centimeters into the sediment of the seafloor.

The T-sensors sampled at a rate of 0.5 Hz. Below floatation at $z = -2310$ m, the extended central line held a single-point 2-MHz Nortek AquaDopp current meter (CM) sampling data at a rate of once per 150 s.

NIOZ4 are self-contained high-resolution T-sensors with a precision better than 0.0005 °C, a noise level of less than 0.0001 °C and a drift of about 0.001 °C mo$^{-1}$ after aging of the thermistor electronics (van Haren 2018). Every 4 hours, all T-sensors on the 5-L were synchronized via induction to a single standard clock, so that individual sensor's times were less than 0.02 s off.

**2.2 Mooring site and conditions**

5L was deployed at 42° 47´N, 06° 09´E, seafloor at $z = -2480$ m, about 40 km south of Toulon, France (Fig. 1c). The average local seafloor slope is less than 1º, a relatively flat topography 12 km seaward of the steep continental slope. Mooring deployment was on 18 November 2017 (yearday 321), recovery on 15 September 2018 (yearday 257 + 365 = 622).

The T-sensors performed well, but only for the first 4.5 months after deployment. A bad make of batteries caused 50% failure around day 460 and more after that. Analysis thus focuses on data from days before 450, when less than 49 (15% of the) T-sensors were either not working, showed calibration problems or were too noisy. For these <15% of T-sensors, data are interpolated between vertical neighbour sensors. During the period between days 321 and 450, deep dense-water formation was not noticed in the mooring area.

For calibration purposes and to establish the local temperature-density relationship, shipborne Conductivity-Temperature-Depth (CTD) profiles were obtained about 1 km West from the mooring site during the deployment and recovery cruises. The 2017CTD was stopped at $z = -2475$ m. The 2018CTD was stopped at $z = -2400$ m due to winch constraints. In October 2020, an additional CTD-profile was obtained to $z = -2479.5$ m at 6 km Northeast of the mooring site.

## 3 Observations

Before discussing analysed T-sensor data in some detail, shipborne CTD-observations and general time-series observations are presented below. In hindsight, CTD-observations also occasionally show about 130-m high unstable waters extending above the seafloor. Following the time-series from the moored T-sensors, convection-turbulence instabilities are best observable in early winter (van Haren 2023).

### 3.1 CTD-profiles and temperature-density consistency

Between the three autumnal CTD-profiles similar, albeit not identical, vertical variations are seen on the large scale in pressure-corrected Conservative Temperature (IOC et al. 2010) $\Theta$ (Fig. 2a), Absolute Salinity (Fig. 2b) and in density anomaly $\sigma_2$ referenced to a pressure level of $2\times10^7$ N m$^{-2}$ (Fig. 2c). The vertical density stratification, which corresponds to -d$\sigma_2$/dz $\propto$ N$^2$, steadily reduces from near the sea-surface towards the seafloor.

In the lower 300 m above the seafloor, the CTD-profiles do vary, not only on small 10-m vertical scales, but also on large 100-m vertical scales (Fig. 2d-f). The 2020CTD-profile shows less small-scale variations than the other two profiles, while most eye-catching is the 'counter-gradient' increase of $\Theta$ with depth (-z) for the 2017CTD-profile (Fig. 2d). A counter-gradient $\Theta$(z)-profile would imply an unstable layer of some 300 m, at least. However, it is (over-)compensated by the large-scale salinity-gradient (Fig. 2e), so that most of the density-profile is stable (Fig. 2f).

Careful inspection shows however, that the lower 130 m above the seafloor are unstable on the large-scale in $\sigma_2$(z) of 2017CTD. As a result, for the range of moored T-sensors the temperature-salinity (Fig. 3a), and the temperature-density (Fig. 3b), relationships are consistent between the three CTD-profiles. As a result, temperature can be taken as a tracer for density variations with salinity contributions included, and temperature profile inversions can be used to quantify turbulence values (Appendix A). It is seen in Fig. 3a that the governing trend in the temperature-salinity relationship is about perpendicular to the isopycnals (contours of constant density), which points at diapycnal transport in this vertical range of 130 m above the seafloor.

For z < -2250 m, stratification is very weak as N calculated from CTD varies between 0 and 2f, with a general 100-m-scale mean value of about $N = 1f = 1.36$ cpd (short for cycle per day), as may be inferred from comparing vertical slopes in Fig. 2f. In the same vertical range, maxima of small-1-m-scale buoyancy frequency $N_s$ are found up to about $N_{s,max} = 4f$. The vertical range of N = 0 (homogenous, neutral conditions, no stratification) is <100 m from the seafloor in these profiles.

**3.2 General winter conditions**

Time series from moored CM-data show that flow speeds were typically <0.1 m s$^{-1}$ in autumn and early winter and doubled occasionally with peaks of 0.35 m s$^{-1}$ due to increased eddy-activity in late-winter (Fig. 4a). One-third of the temperature variations is attributable to instrumental drift causing a low-frequency nonlinear increase with time, mainly during the first month (Figure 4c). Occasional short-term negative temperature differences are observed especially between days 350 and 400. In late winter after day 430, more positive temperature anomalies are found. Some of the latter peaks associate with stronger water-flows, and with acoustic echo amplitude increases (Fig. 4b). The 2-MHz acoustic amplitude reflects variations in suspended particles and small, about 1-mm size zooplankton density, but in a highly qualitative manner as the instrumentation has not been calibrated with local plankton-net data. Nonetheless, the gradual wintertime increase is obvious.

For the purpose of starting the investigation of potential stable and unstable conditions, two vertical temperature difference $\Delta\Theta$-records were computed over large O(100) m and medium O(10) m scales (Fig. 4d). Late-winter positive $\Delta\Theta$-peaks reflecting stable stratification stand out, not only over $\Delta z = 98$ m but also over 16-m vertical scale near the seafloor. These peaks associate with the warm-water peaks in Fig. 4c. Positive ΔΘ imply very small homogeneous layers above the seafloor, which are generally <100 m and occasionally <16 m during the late-winter period.

Both ΔΘ-records also show significant negative values that reflect unstable conditions. Largest negative ΔΘ are observed in early winter (between days 350 and 400), well before positive peaks occur generally (Fig. 4d). Negative ΔΘ indicate rather persistent large-scale instabilities, possibly related to convection-turbulence. Observed ΔΘ = -0.001 °C over 98 m and stable N = 2f higher-up (Fig. 2f) translate to a net vertical homogeneous range of 250 m from the seafloor, which well exceeds the 109-m range of the moored T-sensors. Such inferred near-homogeneous layer above the seafloor is about the largest for the entire record.

**3.3 Homogeneous columns**

Although 250-m large vertically homogeneous waters have not been observed by the 109-m tall mooring, occasional direct observations of large homogeneities are observed to last longer than 2000 s. Such data are necessary for important referencing in corrections of instrumental (electronic) drift (van Haren 2022).

For proper drift correction not resulting in unrealistic unstable conditions, portions of data are helpful in which dΘ/dz = 0 < 0.00001°C/100 m over the entire range of observations as in homogeneous waters providing neutral conditions. Such homogeneous waters exist in the deep Western Mediterranean as established from previous extensive CTD-observations, especially in the central and southern parts of the basin, but these extend up to about 800 m above the seafloor (e.g., van Haren et al. 2014). This height above the seafloor is two-thirds of the vertical range of 1200 m suggested by Ferron et al. (2017). Closer to the continental slope, smaller homogeneous layers of <250-m vertical extent (Fig. 2f) than in the open basin are due to extensive (sub-)mesoscale activity imposing stratification from above related with the boundary flow and later-in-time occurrence of deep dense water formation (Albérola et al. 1995; van Haren and Millot 2003).

In the present T-sensor records, a few periods, with a minimum of 30-minutes duration for statistical reasons, were identified in which temperature variations over 100-m vertical extent were not significantly different from NIOZ4 noise limits, see for example line-c data in Fig. 5. The mean smooth pressure-(adiabatic lapse rate)-corrected zero-slope temperature 'reference' profile is used to replace the mean drift-affected values for each T-sensor in that period. The zero-slope (zero-order polynomial constant fit) smooth profile is referenced to local CTD-data, for absolute accuracy values. As the drift is time-dependent, for periods away from the zero-slope period best-fit polynomials are computed of orders that depend on the standard deviation of the drift with respect to the instrumental noise level. The standard deviation is measured from the mean peak-to-peak drift data for the analysis period.

The relatively warm >100-m high column between days 439.0 and 439.15 in Fig. 5a may have a width of about 2 km, assuming advection by observed 0.15 m s$^{-1}$ mean flow. Unknown is its height, but outside its approximately 400-m wide (2700-s duration) homogeneous core with zero mean profile (Fig. 5b) it shows stable stratification, albeit appearing in thin, rugged columns from above. Around the large column, waters are unstable with highest temperatures observed at (5 m above) the seafloor and thin vertical columnar motions that reflect convection-turbulence (Dalziel et al. 2008).

Highest temperatures are observed at 0.5 m above the seafloor in corner-line data during various occasions of convection-turbulence in the record between early- and late-winter (Fig. 6). The detailed magnifications of $\Theta(t, z)$ show the blooming and plumes of convection-turbulence, not so much occurring in thin columns as well as in highly variable motions distributed over many scales. Like in clouds and bush fires. Given the observed $\Delta\Theta = 0.0005°C$, both Nu $\approx$ 3000 and Ra $\approx 7\times10^9$ indicate convection-turbulence, for typical L = 10 m. For 100 mW m$^{-2}$ geothermal heating and unbounded fluid in high (turbulent) Ra, time between convection plumes is calculated to be about 2200 s using formulation in (Foster 1971; Thorpe 2005). This time between plumes, of about 0.025 day, corresponds well with the intermittency of observations in Fig. 6.

Considering the average advection speed of 0.04 m s$^{-1}$, a plume-structure reaching 100 m above the seafloor would take 0.03 days to pass the mooring. For example, the plume between days 362.31 and 362.34 in Fig. 6a thus represents a quasi-isotropic turbulence motion with approximately equal horizontal and vertical scales, assuming Taylor's hypothesis of frozen turbulence is valid here. This hypothesis allows for a transfer between spectral wavenumber and frequency spaces. While the provided examples have varying intensity and duration, the longest exceeds 1 day, i.e., well longer than one inertial period (Fig. 4d). As a result, it cannot represent shear-turbulence. Indeed, the detailed observations lack any primary roll-up or roll-over imaging that is typical for KHi (Smyth and Moum 2012). Spectral information below distinguishes convection- from shear-turbulence, also for other periods of observations.

**3.4 Temperature spectra**

Frequency ($\sigma$) spectra of temperature variance are observed to be featureless without peaks (Fig. 7). For interpretation in terms of wavenumber spectra, Taylor hypothesis of frozen-turbulence approximation may be invoked (Tennekes and Lumley 1972; Thorpe 2005). The upper and lower T-sensor spectra are scaled with (divided by) $\sigma^{-5/3}$ which is indicative of an inertial subrange or passive scalar shear-turbulence (Ozmidov 1965; Tennekes and Lumley 1972; Warhaft 2000). In the relatively narrow non-traditional internal wave band f > $\sigma_{min}$ < $\sigma$ < $\sigma_{max}$ > N for meridional propagation (LeBlond and Mysak 1978), the spectra generally align with a slope of +2/3 in the log-log plot, i.e., align with $\sigma^{2/3}$ which is equivalent to align with $\sigma^{-1}$ in an unscaled plot. The $\sigma^{-1}$-aligning has previously been observed in broader-band, larger stratification open-ocean data (van Haren and Gostiaux 2009). It also reflects pink-noise intermittency and is broadly seen as the generation of barely stable structures of critical states (Schuster 1984; Schroeder 1991). Between about $\sigma$ = 2.5f and roll-off frequency 300 cpd, the upper T-sensor spectra align with zero slope and thus (in an unscaled plot) with $\sigma^{-5/3}$ of shear-turbulence inertial subrange between the internal wave band and (unresolved) turbulence dissipation.

The upper-T-sensor spectral-slope alignment with the inertial subrange slope is observed to be dominant throughout the 3.5-month record (Fig. 7a), and in all particular 18-d sub-periods of: early-winter dominant N=2f stratified conditions with relatively weak upper T-sensor variance and relatively soon roll-off to noise at about 100 cpd (Fig. 7b), early-winter dominant unstable (presumably geothermal) conditions (Fig. 7c), and late-winter mixture of conditions of unstable geothermal and stable N=4f≈$N_{s,max}$ internal-wave turbulence from above with relatively large upper T-sensor variance and thus spectral extent to about 800 cpd before rolling off to noise (Fig. 7d).

In contrast, in all panels of Fig. 7 for about $N_{s,max} < \sigma < 200$ cpd the lower T-sensor spectra demonstrate little aligning with the inertial subrange. Instead, a statistically significant (unscaled-plot) aligning is observed with $\sigma^{-7/5}$ that reflects an active scalar (Bolgiano 1959; Pawar and Arakeri 2016) and convection-turbulence. Between the frequency ranges of $\sigma^{-1}$- and $\sigma^{-7/5}$-alignings, the lower T-sensor spectra align with $\sigma^{-2}$ or steeper over a range that seems to depend on the stratification: from the frequency of the highest T-variance related to buoyancy (between [N, $N_{s,max}$]) and twice that frequency. Thus, between [2f, 4f] in Fig. 7b (and curiously Fig. 7c) and between [4f, 8f] in Fig. 7d. An aligning with $\sigma^{-2}$ is associated with internal waves (Garrett and Munk 1972) but seems to be partially related to sufficiently intense fine-structure contamination as well here (Phillips 1971). During some periods the $\sigma^{-2}$-aligning is also observed in upper T-sensor spectra (Fig. 7b, d). The largest dip or discrepancy between lower T-sensor spectra and upper T-sensor spectra is found around 10 cpd. Although this dip does not reflect a spectral gap in an unscaled plot, the apparent lack of inertial subrange and thus shear-turbulence in lower T-sensor spectra suggests different processes of energy transfer between internal waves and convection near the seafloor. Such transfer may not occur locally.

Thus spectrally, T-sensors show a limited internal-wave band related with the weak stratification throughout the vertical range of observations and two orders of magnitude extended inertial subrange of shear-turbulence away from the seafloor, with convection-turbulence activity being dominant closer to the seafloor. Variance is generally significantly larger for upper than for lower T-sensors.

**3.5 Transitions from anisotropic to isotropic turbulence**

5-L has 260 T-sensors that can be used for vertical and horizontal statistics. For each of the 52 overlapping vertical positions, coherence (coh) can be computed between records from 2-4 independent T-sensor pairs across horizontal $\Delta x,y$ = 4, 5.6 and 8 m. The statistical average is compared with vertical coherence between all possible pairs of independent T-sensor records across vertical $\Delta z$ = 2, 4 and 8 m. Per frequency, the interpretation of anisotropic (stratified) turbulence follows when the horizontal coherence is larger than the vertical for the same (horizontal, vertical) distance between sensors. Isotropic turbulence follows when the coherence level is identical for the same distance.

The above method works well when the large-scale 'advective' water-flow is negligible. However, depending on length-scale, atmospheric boundary layer turbulence shows different temperature variance in along-wind and across-wind directions as has been demonstrated from aircraft observations (Nicholls and Readings, 1981). At large (turbulence inertial subrange) length-scales along-wind variance was larger than across-wind, and vice versa at small length-scales. This length-scale dependence was found to be consistent with stretching effects of the mean velocity shear acting on convective elements. Although such horizontal anisotropy has not been investigated previously for ocean turbulence to the knowledge of the author, the present moored T-sensor data show a tendency for higher coherence in crossflow direction compared to alongflow direction. The coherence for the alongflow direction is thus considered low-biased by the analysis scheme, by an amount of Δcoh = 0-0.05 depending on the turbulence frequency. This low-bias for alongflow direction has been established by investigating coherence for short data records with steady (constant amplitude) flows in either x- or y-direction of the rigid mooring.

For the present analysis, horizontal coherence data are retained that are collected in crossflow direction to within ±25°, i.e., in North-South direction for the examples given below with dominant East-West flows and adopting validity of Taylor's hypothesis. As a criterion for checking, it is assumed that horizontal coherence cannot be smaller than vertical coherence for given distancing, commensurate the suppression of vertical turbulence scales by stratification.

Two short periods of geothermal convection of Fig. 6a and 6d are investigated for coherence in Fig. 8. Temperature variance is given for reference in Fig. 8a, c and for comparison with Fig. 7 (albeit over a reduced frequency range that focuses on the turbulence portion of data.) In both examples the water-flow speed averaged 0.05±0.01 m s$^{-1}$.

Fig. 8a, b shows an example of convection-turbulence near the seafloor with, spectrally, shear-turbulence with inertial subrange aligning for the upper T-sensors (Fig. 8a). Across the one-order of magnitude large frequency range at $\sigma$ < 300 cpd before roll-off to instrumental noise, coherence across small <10-m scales is seen to transit from highly significant coh > 0.8 values to insignificant noise levels coh ≈ 0.15 (Fig. 8b). Within this transition range of coherence-values, the vertical:horizontal distance ratio $c_r = \Delta z:\Delta x,y$ of equal coherence slowly drops from about $c_r$ = 0.5 to 1, and isotropy ($c_r$ = 1) is reached at about 300 cpd. The relatively large range of dominant anisotropic turbulence is attributed to the effects of shear-turbulence at upper T-sensors, which apparently provides a dampening of convection-turbulence that is limited to <100 m from the seafloor. Such dampening is difficult to infer visually from the time-depth image (Fig. 6a). In early winter, stratification apparently still has effects on (the direct observation of) geothermal heating, thereby limiting its vertical extent.

Fig. 8c, d shows a contrasting example of intense convection-turbulence and about equal variance for all T-sensors (Fig. 8c). Because of the lower temperature variance in Fig. 8c compared to Fig. 8a, the one-order of magnitude large transition of coherence before roll-off to instrumental noise in Fig. 8d shifts to slightly lower frequencies compared to Fig. 8b. Throughout the transition range in Fig. 8d, isotropic $c_r = 1$ for 4-m separation distance, while for 8-m distance isotropy is reached for $\sigma > 80$ cpd. The shorter transition range of dominant anisotropic turbulence, compared to Fig. 8b, reflects convection-turbulence extending above the 108-m range of T-sensors. While such vertical convection-process is also difficult to infer visually from the time-depth image (Fig. 6d), it occurs around the time when stratified-convection is occasionally accelerated from above. That is, just before potentially new stratification is advected into the area.

The coherence spectra suggest that convection-turbulence provides isotropy at lower frequencies than shear-turbulence, with lower coherence-values at higher frequencies for the former compared to the latter. Over time from early to late winter, the geothermal reduction of stratification is less compensated by restratification, and isotropy is reached faster and reaches higher, although only about 100-250 m, from the seafloor.

### 3.6 The turbulence passband

The band in which the short-scale coherence drops from about 0.9 to levels of insignificant values, including the transition from anisotropic stratified turbulence to isotropic turbulence, is separated from the T-sensor signals for further investigation. The data of Fig. 6 are band-pass filtered (bpf) between [20, 200] cpd using sharp phase-preserving double-elliptic filters (Parks and Burrus 1987). This [20, 200]-cpd band is the frequency range, well outside the internal wave band, in which turbulence shows maximum variance-discrepancy between upper and lower T-sensor spectra (Figs 7, 8a), if convection does not cover the entire 108-m T-sensor range (Fig. 8c). The result is shown in Fig. 9, with a common total temperature range of 0.0003°C. The bpf-frequency range describes the larger turbulence scales, which are either predominantly shear-turbulence and fill the spectrum between the internal wave band and instrumental noise (Fig. 9a, b for the upper T-sensors mainly), or convection-turbulence (Fig. 9d).

In case of shear-turbulence dominance, the lack thereof in the lower T-sensors is visible in relatively low variance (grey colour near the seafloor in Fig. 9a, b). The lack of, or weak, shear is found in a layer of maximum 10 m above the seafloor, which varies in height over time. Shear-turbulence in the upper T-sensors does not show KHi-roll-up but is more associated with erratic convection. The shear is presumably of a secondary (smaller) type along the edges of convection (Li and Li 2006), which in the mean dominates over primary convection.

More columnar convection-turbulence is visible in Fig. 9c, d. A (one-)inertial period variation with time of columnar intensity is noted in Fig. 9d. The columns occasionally reach to (within 0.5 m from) the seafloor. Seldom, a column extends from the seafloor up to 109 m above it without loss of intensity. Generally, columns are slanted, broken and otherwise variable in the t,z-plane.

In all examples, the order of most intense bpf-variance is in sequence of Fig. 9-panels b, a, c, d (see also Table 1). The b,a-sequence is identical to one on the overall size of temperature-ranges of Fig. 6. As that figure is dominated by the lower frequency motions of the internal wave band, the consistency of temperature ranges suggests a distinctive link between internal waves and turbulence. While the link is likely downscale for shear-turbulence following the general Kolmogorov-Ozmidov (Kolmogorov 1941; Ozmidov 1965) theory-model of a forward energy cascade, the link may be upscale for convection generating internal waves, as has been demonstrated for a model stratification above a boundary layer that is heated from below (Michaelian et al. 2002). Their model shows transfer of convection energy to internal waves and mean flow, and, over time, organization of convection-plumes that are initially irregular.

The standard deviation (std) of the band-pass filtered signals of Fig. 9 demonstrates a consistent image of large variability and generally higher values away from the seafloor, mostly lowest values in the lower 10 m above the seafloor with a minimum at $z = -2477.5$ m (Fig. 10), except for green (panel-d) curves when most vertical plumes and lowest T-variance were observed. The slightly increasing values in the 2.5 m towards the seafloor may reflect conduction of the sediment-temperature which has a typical gradient of -0.1°C m$^{-1}$ (e.g., Louden et al. 1997; Pfender and Villinger 2002). Also consistently between the independent T-sensors from the four corner-lines is reduced temperature variability between the lowest two sensors. It is noted that only at the low std-level of about $2\times10^{-5}$ °C convection-turbulence spectral slope has been observed, generally at lower T-sensors within 10 m from the seafloor, and specifically in the late-winter example of Fig. 9d over the entire T-sensor range.

The 10-m distance from the seafloor in which the low std-level of bpf-turbulence temperature variance is observed during all examples seems to indicate a separation of turbulence-convection development. In numerical experiments of near-surface convection by Julien et al. (1996) it is visible in the rotational case. Although it is observed above a flat seafloor with negligible topographic features and general slope angle, it is not directly associated with a 'boundary layer'. In general terms, a boundary layer is understood to result from a linear frictional flow over a flat plate, and in geophysical flows on a rotating Earth it is described as a balance between pressure gradient force, Coriolis force and turbulent drag. The resulting 'Ekman-layer' height of frictional influence controlled by rotation measures $\delta = \sqrt{(2A/f)}$, in which A denotes the turbulent viscosity that is tacitly assumed to result from frictional shear-induced turbulence.

Taking $\delta = 10$ m we find $A \approx 5\times10^{-3}$ m$^2$s$^{-1}$. However, noting that this 10-m seafloor distance is inferred from turbulence convection, one may question whether we deal with a 'normal Prandtl' boundary layer leading to shorter 'turbulent mixing lengths' (Tennekes and Lumley 1972). As will be discussed below, one may consider more generally shear- or convection-turbulent eddies to be dampened near the flat seafloor boundary.

**4 Discussion**

For years, geophysicists have established general geothermal heating in the deep-sea via heat-flow measurements of a combination of temperature gradients and thermal conductivity in sediments. The results are presented in global maps of Earth heat-flow with typical flux-values of 100 mW m$^{-2}$ for the Western Mediterranean (Pasquale et al. 1996; Davies and Davies 2010). Modelling the effects of this general geothermal heating on the development of temperature (density) variations in deep waters and overturning circulation suggested that about 25% of the heating is used for turbulent (convective) diffusion (Adcroft et al. 2001). Most of the heat is transported via advective currents. From their box-model of the ocean overturning circulation, Mullarney et al. (2006) concluded that the destabilizing geothermal heat flux promotes a more vigorous overturning having approximately 10% greater volume flux than without seafloor heating. No significant change was seen by them in the vertical density structure, which is not surprising considering the difficulty in observing these in the deep-sea under near-homogeneous conditions, and the warming (stabilization) from above.

Laboratory works and Direct Numerical Simulation (DNS) modelling have demonstrated the details of Rayleigh-Bénard and RTi turbulent convection, also in a rotating frame of reference (e.g., Bartello et al. 1994). When rotation dominates over buoyancy, turbulence convection organizes in near-vertical tubes of up- and down-going waters (Davidson et al. 2006; Staplehurst et al. 2008). The organization in near-vertical tubes should occur within a timescale of 1/(rotation frequency). It is shown in the present observations that started within 0.5 m from the seafloor that the tubes become thinner 10-100 m away from the heating source, whereas temperatures are more uniform with reduced tube-formation in the lower 10 m above the seafloor. While the individual plumes have a much shorter timescale than the inertial period $2\pi/f$, and even than $1/f$, interaction is expected with (sub-)mesoscale eddies. Such eddies include inertial (super-)harmonic motions, which orient to the Earth rotational vector $\Omega$ resulting in very weak N=O(f) stratification (Straneo et al. 2002), but it is unclear if orientation of $\Omega$ matches the slanting of plumes observed in Fig. 6.

Because of the condition of very weak stratification, direct observation of unstable general geothermal heating requires moored high-resolution sensors that can resolve 0.00005 °C over ranges of typically only 0.0004 °C. The condition of weakly stratified waters should remain so without strong restratification processes such as advection, internal waves and their turbulent breaking and/or heating from above. Reformulating Adcroft et al. (2001) on the percentage of geothermal heating going into turbulent diffusion, the condition of very weak stratification for direct observation of convection occurs 25% of time. This percentage corresponds with the occurrence of 24% of ΔΘ-instabilities over 98-m range in Fig. 4d using a threshold of 2x0.00005 °C.

Under weakly stratified conditions, the Western Mediterranean moored T-sensor and shipborne CTD-observations near the continental slope demonstrate general geothermal heating that extends about 50-150 m above the flat seafloor, mostly in early-mid winter. This vertical extent is considerable, but only one-tenth of the value anticipated in the open basin (Ferron et al. 2017). Spectra from moored T-sensor observations demonstrate turbulence subrange slopes of dominant shear-turbulence evidencing non-negligible stratification near the top of the 109-m tall mooring in most cases. Dominant convection-turbulence is often observed in spectral slopes, mostly from the lower 50 m above the seafloor. Only occasionally, such buoyancy subrange slopes are observed across the entire 109-m range of observations.

Despite the expected secondary shear-instabilities along the edges of convection-plumes, no clear spectral evidence is found for a transition from convection- to (an inertial subrange of) shear-turbulence, except perhaps a small indication thereof around 100 cpd in Fig. 8c in the upper T-sensor data mainly. The start of the inertial subrange is approximately between 10-30 cpd. The variability in precise transition frequency to $\sigma^{-5/3}$-aligning for convection-turbulence waters (Figs 7, 8a, c) requires further investigation. On first thoughts, the transition to inertial subrange is unlikely related with the Ozmidov frequency of largest isotropic turbulent scales, because the Ozmidov scale is generally assumed to hold for turbulence in (near-homogeneous layers in) well-stratified waters, not for convection-turbulence.

However, in contrast with (laboratory) convection-turbulence of RTi across the entire fluid domain between surface and bottom, the Western Mediterranean convection-turbulence near the seafloor is capped by stable stratification above. Although such stratification is often outside the range of the present moored T-sensors, it is visible in CTD-profiles. The capping by stratification is reminiscent of stable stratification underneath a layer of (nighttime) convection turbulence near the sea-surface. For such near-surface area, turbulence values have been successfully computed by Kumar et al. (2021) using the method of reordering observed density profiles that may contain unstable portions and comparing the displacements between observed and reordered profiles with the Ozmidov scale. The method-use of reordering seems common sense, even though the method was originally proposed for shipborne CTD-observations in generally well-stratified lake-waters (Thorpe 1977). As long as there is stratification capping an unstable layer above or below, or both, there seems no difference in the assumption that the turbulence process works against the stable stratification that follows after the reordering of the (partially) unstable profiles in stable ones, as is verified below for geothermal convection observations using the moored T-sensors.

For the short convection periods of Fig. 6 the root-mean-square Ozmidov scale amounts $L_O = 45\pm10$ m using the standard computation (A3) as given in Appendix A. While this $L_O$-value may reflect the largest overturn sizes in Fig. 6, see also Fig. 10 for the vertical extents of large and reduced turbulence temperature variance also towards the seafloor, its corresponding Ozmidov transition frequency $U/2\pi L_O = 4.5\pm1.5$ cpd for average $U = 0.1$ m s$^{-1}$ seems too low to indicate the transition to turbulence inertial subrange, although it is within one std from mean $N_{s,max}$ (Table 1), the presumed limit of internal waves.

Alternatively, one may consider the small-scale stratified layers and assume that the (secondary shear-) turbulence works on these, so that convection must overcome the strongest small-scale stratification that, for example, is found in thin layers adjacent to convection-plumes (Li and Li 2006). For each vertical profile of T-sensor data, it is possible to determine the maximum of small-scale buoyancy frequency $N_{s,max}(t)$ and replace N in (A3) to compute $L_{Os} = 17\pm5$ m and $U/2\pi L_{Os} = \sigma_{Os} = 13\pm4$ cpd. The 'small-scale' Ozmidov frequency $\sigma_{Os}$ is commensurate the observed range of transition to inertial subrange, and the (low-frequency) start of transition from anisotropic to isotropic turbulence. Arguably, it is a better indicator than 10 m of the range above the seafloor across which turbulence $\Delta\Theta$-variance is reduced (Fig. 10). In terms of convection, it would indicate the layer of near-uniform warmer waters over the seafloor across which plumes have limited effect. As was suggested in van Haren (2023), the above separation in (stratification) scales may also be useful in turbulence value computations.

Computing the turbulence dissipation rate following conventional method (A2) gives average values of about $2\times10^{-8}$ m$^2$ s$^{-3}$ for the convection-turbulence periods in Fig. 6 (Table 1). It is assumed that these short periods contain negligible restratification, e.g., via advection, over the range of observations. Vertical integration of the average turbulence kinetic energy dissipation rate over the 109-m observational range provides a value of about 2.2 mW m$^{-2}$ after conversion with density of seawater $\rho$. This $\int\rho\varepsilon dz$-value is just 2% of the average amount of the vertical flux attributed to geothermal heating from heat-flow measurements in the area (Pasquale et al. 1996). If a turbulent mixing efficiency, proportion of kinetic energy consumed by mixing, is considered of 0.5 which is typical for vertical natural convection (Dalziel et al. 2008; Gayen et al. 2013; Ng et al. 2016), the calculated integral dissipation rates here fall short by more than one order of magnitude of a presumed value of 50 mW m$^{-2}$. To compensate one requires integration over at least 2500 m from the seafloor. Given general suppression of turbulent overturning in stratified waters that have been observed by CTD above, such 'entire water column' integration is likely to yield (considerably) smaller results and thus a discrepancy with the Earth heat-flow flux. Two suggestions to explain this discrepancy were given in van Haren (2023). They are elaborated below with application to the observations in Fig. 6.

First, adopting the same reasoning for small-scales affecting the Ozmidov scale, one may, for each vertical profile of T-sensor data, replace N in (A2) by $N_{s,max}(t)$ to compute 'small-scale' dissipation rate,

$$\varepsilon_s = 0.64 d^2 N_{s,max}^3. \qquad (1)$$

The mean value of dissipation rate $\varepsilon_s$ (in $m^2 s^{-3}$) gives, after averaging over the short periods of Fig. 6 and after integration over the 109-m vertical range multiplied by $\rho$, 22 mW $m^{-2}$.

Second, from the shipborne CTD-data it is inferred that the vertical density gradient becomes steep enough so that large-scale $N > 2f$ at $z = -2225\pm75$ m, about 250 m above the seafloor. Under the assumption that convection reaches with the same turbulence intensity that far from the seafloor, the integration over 109 m may be extended by $150\pm75$ m for the observations in Fig. 6. The variation in extended integration may (inversely) reflect the variation in dissipation rate for the different short periods (Table 1). In conjunction with (1), 150-m extended integration yields a mean value of $\int\varepsilon_s = 53\pm15$ mW $m^{-2}$. This $\int\rho\varepsilon_s dz$-value corresponds, to within error, with 50% of the geophysically determined heat-flow through the Earth crust (Pasquale et al. 1996).

In combination, the observations and above analysis demonstrate the much smaller vertical extent of convection-turbulence by geothermal heating than computed for the central basin in the Western Mediterranean (Ferron et al. 2017). The smaller vertical extent of convection-turbulence confirms the notion that the present observations are close to an area, the continental slope, which is a source for sufficient restratification to dampen the convection-turbulence, presumed mostly via (sub-)mesoscale eddies.

As for the apparent discrepancy or separation of scales between primary and secondary convection and shear, a remaining issue is the size of length-scales above the seafloor. Judging from the $\Delta\Theta$-convection-turbulence variance in Fig. 10, the 10-m scale of minimum variance above the seafloor corresponds with a turbulent viscosity which is one order of magnitude smaller than the mean turbulent diffusivity in Table 1. However, as turbulence is not a property of the material but of the flow (Tennekes and Lumley 1972), it is unlikely that the two turbulent exchange coefficients can be that much different, and their ratio is expected to be O(1). Thus, assuming $A = K_z = 7\times10^{-2}$ m$^2$s$^{-1}$, using conventional mixing efficiency values of $\Gamma = 0.2$ in (A4), one finds a mean Ekman height of $\delta = 38$ m. Using $\Gamma = 0.5$, one gets $\delta = 59$ m. Both $\delta$-values correspond roughly with (large-scale) $L_O$, i.e., $L_O$ is in between these two $\delta$-values. The correspondence between values of $\delta$ and $L_O$ suggests Earth rotational effects on $L_O$, but not on small-scale $L_{Os}$. However, $\delta = 60$ m is (also) obtained using $\Gamma = 0.2$ and replacing in (A4) N by $N_{s,max}(t)$ which is presumed to be mainly (secondary) shear-driven.

The suggested difference in scales affected by different mixing efficiency remains puzzling however, in terms of overall net effect. Although plausible from the perspective of primary and secondary turbulence types, convection- or shear-induced, the different mixing efficiencies require future investigation. As for the Ekman height above the seafloor, which is obviously here not established by <0.1 m s$^{-1}$ speeds of (mainly inertial) frictional flows shearing, convection-turbulence induced by general geothermal heating may have impact on the general transport of suspended materials, once released from the seafloor, by the considerable increase of nearly one order of magnitude in scale-height, and thus reduced main flow-shear.

## 5 Conclusions

1.Using moored high-resolution temperature sensors, wintertime general geothermal heating has been observed generating primary convection-turbulence above a deep seafloor over short periods O(1-10) h, occasionally lasting >1 d which exceeds the local inertial period. The latter rules out primary generation of stratified turbulence induced by shear due to breaking internal waves, which cannot last longer than the buoyancy and/or inertial period.

2.Periods of convectively unstable temperature profiles in the lower 100 m above the seafloor comprise about 25% of the four months between November and March of moored observations at the deep Western Mediterranean site.

3.The lowest 10-15 m from the seafloor are characterized by reduced turbulent temperature variability, which suggests that turbulence scales are not fully developed.

4.Spectral information demonstrates that in the lower 50 m above the seafloor convection-turbulence is dominant, while higher-up (and occasionally in the lowest meters above the seafloor) geothermal heating associates with (secondary) shear-turbulence.

5.Turbulence values are calculated for periods of dominant convection-turbulence capped by stable stratification using the conventional method of reordering observed density (temperature) profiles into stable ones. The method underestimates the turbulence dissipation rate in waters just above the seafloor by geothermal heat flux. Only when it is assumed that turbulent overturning must overcome smallest stratification scales, good correspondence is found between historic geophysical heat-flow measurements and the present oceanographic water turbulence dissipation rates.

**Acknowledgments** I thank the captain and crew of the R/V l'Atalante and NIOZ-NMF for their very helpful assistance during deployment and recovery and for the construction of the mooring array. I thank M. Stastna (Univ. Waterloo, Canada) for providing the 'darkjet' colour-map suited for T-sensor data.

**Funding** This research was supported in part by NWO, the Netherlands Organization for the advancement of science.

**Data availability** Data that support the findings of this study are available from the corresponding author, upon reasonable request.

**Declarations**

**Conflict of interest** The author declares no competing interests.

**Appendix A Moored T-sensor turbulence values**

Over the vertical range of moored T-sensors the Conservative Temperature-density anomaly ($\Theta$-$\sigma_2$) consistent relationship amounts,

$$\delta\sigma_2/\delta\Theta = -0.85\pm0.05 \text{ kg m}^{-3} \text{ °C}^{-1}. \tag{A1}$$

The relatively tight relationship (A1) implies the T-sensor data may be used as a proxy for density variations and in which salinity contributions are implicitly incorporated. The relationship is useful for inferring turbulence values using the method of reordering unstable data-points to monotonously stable vertical profiles (Thorpe, 1977). Turbulent overturns follow reordering every 2 s the 109-m high (for corner lines) potential density profile $\sigma_2(z)$, which may contain inversions, into a stable monotonic profile $\sigma_2(z_s)$ without inversions. After comparing observed and reordered profiles, displacements $d = \min(|z-z_s|)\cdot\text{sgn}(z-z_s)$ are calculated necessary for generating the reordered stable profile. Then the turbulence kinetic energy dissipation rate reads,

$$\varepsilon = 0.64 d^2 N^3, \tag{A2}$$

where buoyancy frequency $N$ is computed from each of the reordered, essentially statically stable, vertical density profiles.

The numerical constant follows from empirically relating the root-mean-square (rms) overturning scale $d_{rms} = (\Sigma d^2/n)^{0.5}$ over n samples with rms-Ozmidov scale

$$L_O = (\varepsilon/N^3)_{rms} \tag{A3}$$

of largest isotropic turbulence overturns in a stratified fluid as an average over many realizations via the ratio: $L_O/d_{rms} = 0.8$ (Dillon, 1982). This ratio reflects turbulence in any high Reynolds number stably stratified environment like the deep-sea, in which shear-driven and convection-turbulence intermingle at small and large scales and are difficult to separate. In all cases, the mechanical turbulence must work against the stratification that follows from the reordering. It has thus successfully been applied for mainly convection-turbulence (e.g., Chalamalla and Sarkar 2015;

Kumar et al. 2021) while first used for mainly shear-turbulence (Thorpe 1977). Comparison between calculated turbulence values using shear measurements and using Thorpe overturning scales with above constant led to 'consistent results' (Nash et al. 2007).

Likewise, using a constant mixing efficiency of $\Gamma = 0.2$ after substantial and suitable averaging (Osborn 1980; Oakey 1982; Gregg et al. 2018), vertical turbulent diffusivity is computed as,

$K_z = \Gamma \varepsilon N^{-2}$. (A4)

In (A2), and thus (A4), individual d are used rather than taking their rms-value across a single overturn as originally proposed by Thorpe (1977). The reason is that individual overturns cannot easily be distinguished, first, because they are found at various scales with small ones overprinting larger overturns, and second, because some overturns exceed the range of T-sensors. 'Sufficient' averaging is required, also to include various turbulence types of different scales and different age with potentially different $L_O/d_{rms}$-ratio (Chalamalla and Sarkar, 2015) during a turbulent overturn lifetime. While shipborne vertical profiling instruments limit to vertical data averaging, the advantage of a densely instrumented mooring line is also averaging data over time.

**Table 1** Mean values of convection-turbulence for the short observational periods in Fig. 6 and Fig. 9. For all, 105-m large-scale mean buoyancy frequency N = [<N>], in which <> denotes averaging over time and [] over the vertical. The turbulence kinetic energy dissipation rate $\varepsilon$ and vertical turbulence diffusivity $K_z$ are computed using N in (A2) and (A3), respectively. Experimentally, 2-m small-scale per-profile-maximum buoyancy frequency $N_{s,max}$ is used to compute $\varepsilon_s$ as in (1). The standard deviation (std) of band-pass filtered (bpf) temperature refers to Fig. 9.

| Fig.6/9 | $[<\varepsilon>]$ $(m^2\ s^{-3})$ | $[<K_z>]$ $(m^2\ s^{-1})$ | N (cpd) | $[<\varepsilon_s>]$ $(m^2\ s^{-3})$ | $<N_{s,max}>$ (cpd) | std($\Theta_{bpf}$) (°C) |
|---|---|---|---|---|---|---|
| a | $1.5\pm1\times10^{-8}$ | $7\pm3\times10^{-2}$ | 2.2 | $1.8\pm1.1\times10^{-7}$ | 5.9 | $4\times10^{-5}$ |
| b | $4\pm2\times10^{-8}$ | $9\pm4\times10^{-2}$ | 3.2 | $4\pm2\times10^{-7}$ | 7.7 | $6\times10^{-5}$ |
| c | $1.1\pm0.6\times10^{-8}$ | $7\pm3\times10^{-2}$ | 1.9 | $1.1\pm0.6\times10^{-7}$ | 4.8 | $3\times10^{-5}$ |
| d | $8.9\pm0.5\times10^{-9}$ | $7\pm3\times10^{-2}$ | 1.8 | $8.9\pm0.5\times10^{-8}$ | 4.5 | $2\times10^{-5}$ |
| mean | $1.9\pm1.2\times10^{-8}$ | $7\pm3\times10^{-2}$ | 2.3 | $2.0\pm1.2\times10^{-7}$ | 5.8 | $4\times10^{-5}$ |

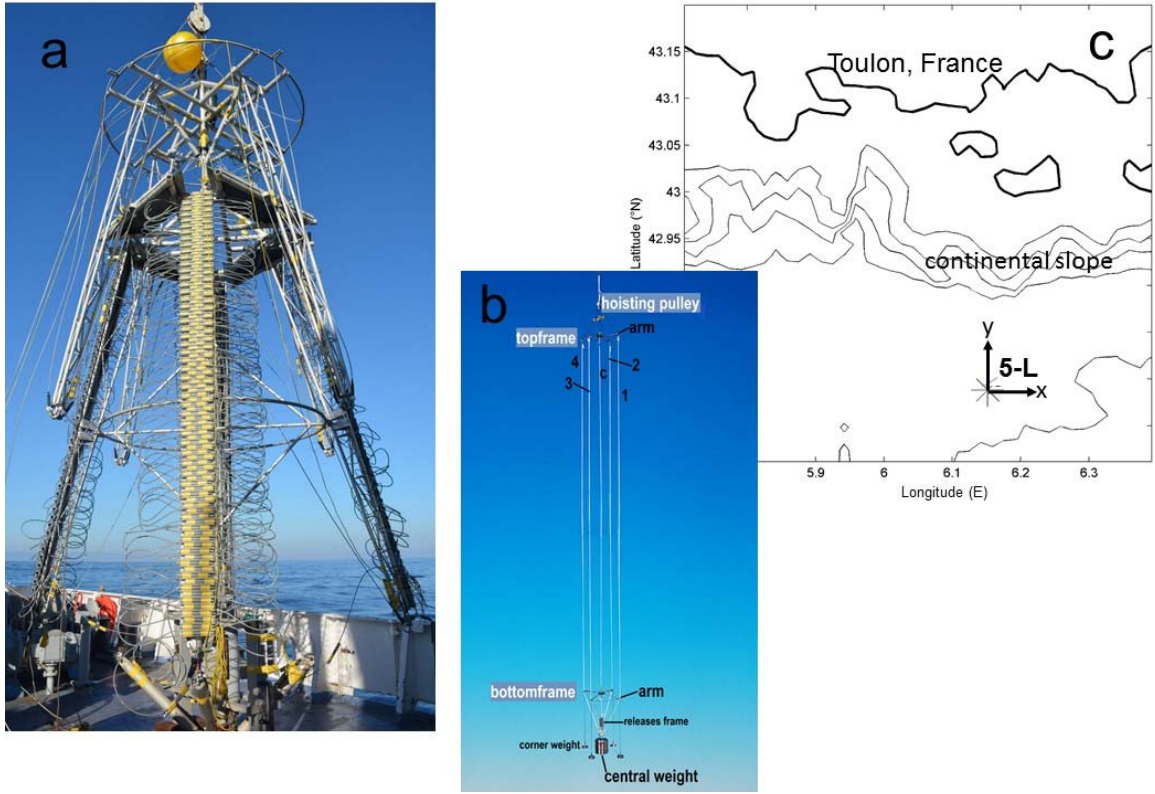

**Fig. 1** Five-line '5-L' mooring construction and site. (a) The 5-L in fold-up form on deck of R/V l'Atalante just prior to deployment. (b) Model of unfolded mooring, to scale. (c) Site of 5-L (star) about 40 km south of Toulon, France, and 12 km south of the foot of the continental slope. Depth-contours are drawn every 500 m.

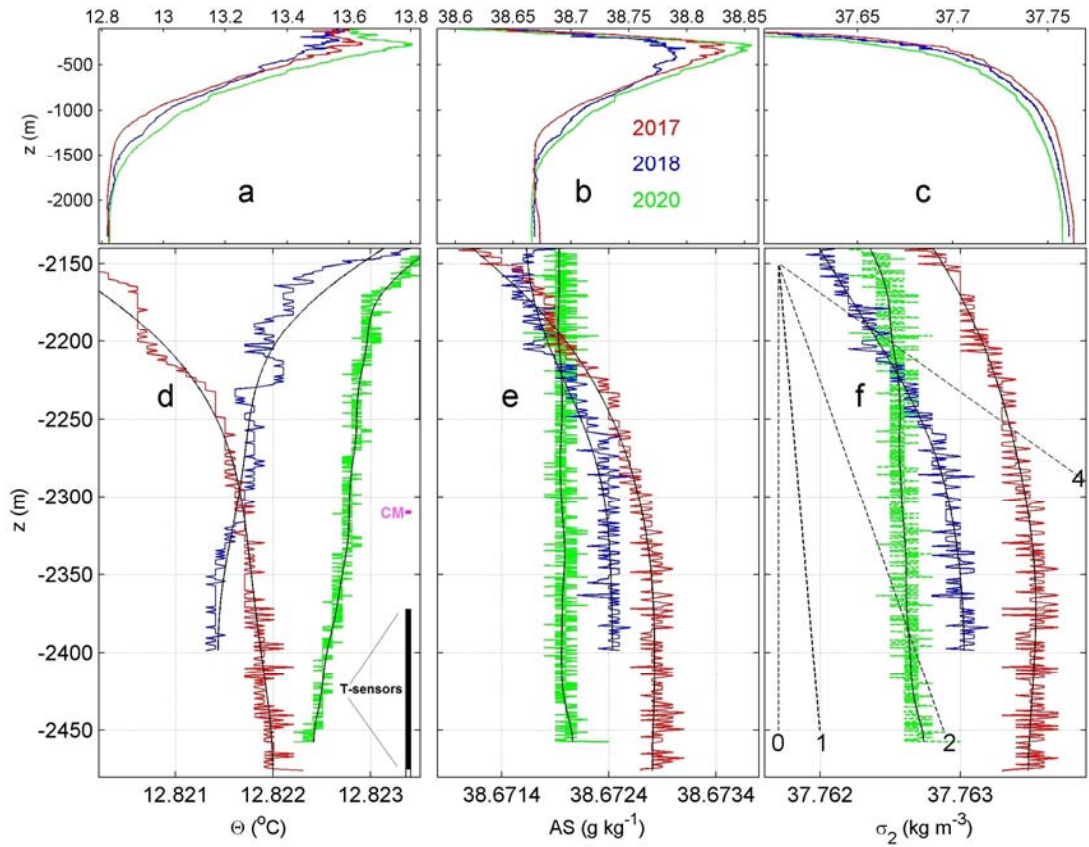

**Fig. 2** Shipborne CTD-profiles obtained near the mooring site during deployment (red; with lowest value 5 m above the local seafloor) and recovery (blue; with lowest value 80 m above the seafloor) cruises with an extra profile from October 2020 (green; with lowest value 0.5 m above the seafloor). Panels a.-c. demonstrate depth-ranges between the seafloor (z = -2480 m) and -100 m. Panels d.-f. show magnifications for -2480 < z < -2140 m, with O(0.001) bias corrections to fit all data within the limited x-axes ranges. Thin black lines indicate low-pass filtered data. (a, d) Conservative Temperature. (b, e) Absolute Salinity. (c, f) Density anomaly referenced to a pressure level of $2\times10^7$ N m$^{-2}$. Panel d. contains a mooring sketch to vertical scale, with CM indicating the current meter. Panel f. contains four vertical slopes (dashed lines) equivalent to buoyancy frequencies N = 0, 1f, 2f, 4f, for reference, where f denotes the local inertial frequency.

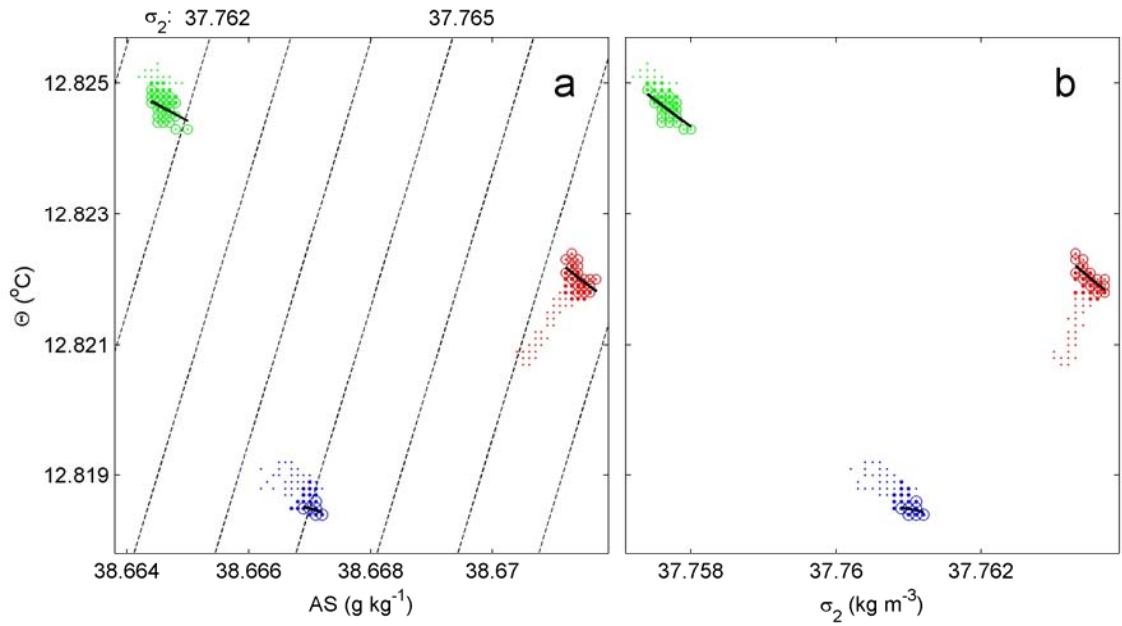

**Fig. 3** Relationship plots for deep CTD-data from Fig. 2d-f, but without bias-correction, using the same colours. Small dots are data between the deepest value of the profile and z = -2200 m. Heavy dots are data between -2400 < z < -2300 m, circles for the range of moored T-sensors between -2480 < z < -2375 m. Short black lines indicate linear best-fits for the 105-m vertical T-sensor range. (a) Conservative Temperature-Absolute Salinity 'TS'-plot. Dashed black density-anomaly contours are drawn every $\Delta\sigma_2 = 0.001$ kg m$^{-3}$. (b) Conservative Temperature-density anomaly plot.

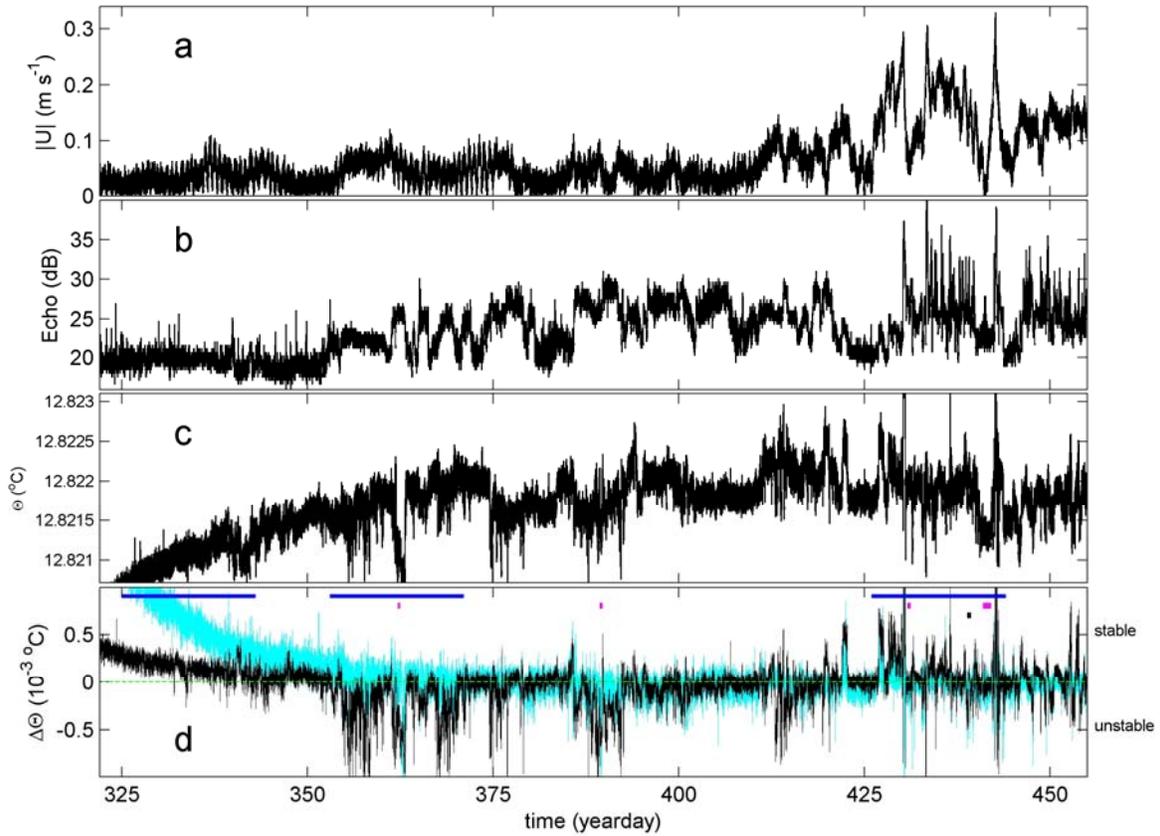

**Fig. 4** Overview of first 4.5 months of moored observations. (a) Water-flow speed at $z = -2310$ m. (b) Acoustical echo amplitude at -2310 m. (c) Conservative Temperature from the T-sensor at -2377 m, uncorrected for bias. (d) Conservative Temperature difference between T-sensors at -2475 and -2459 m (light blue; vertical-axis scale×2), and between -2475 and -2377 m (black). Data are low-pass filtered (lpf) with cut-off at 1000 cpd. The blue bars indicate 18-day periods for which spectra are computed in Fig. 7b-d. The purple ticks indicate short periods for which magnifications are shown in Fig. 6. The black tick indicates the day of reference T-sensor data in Fig. 5.

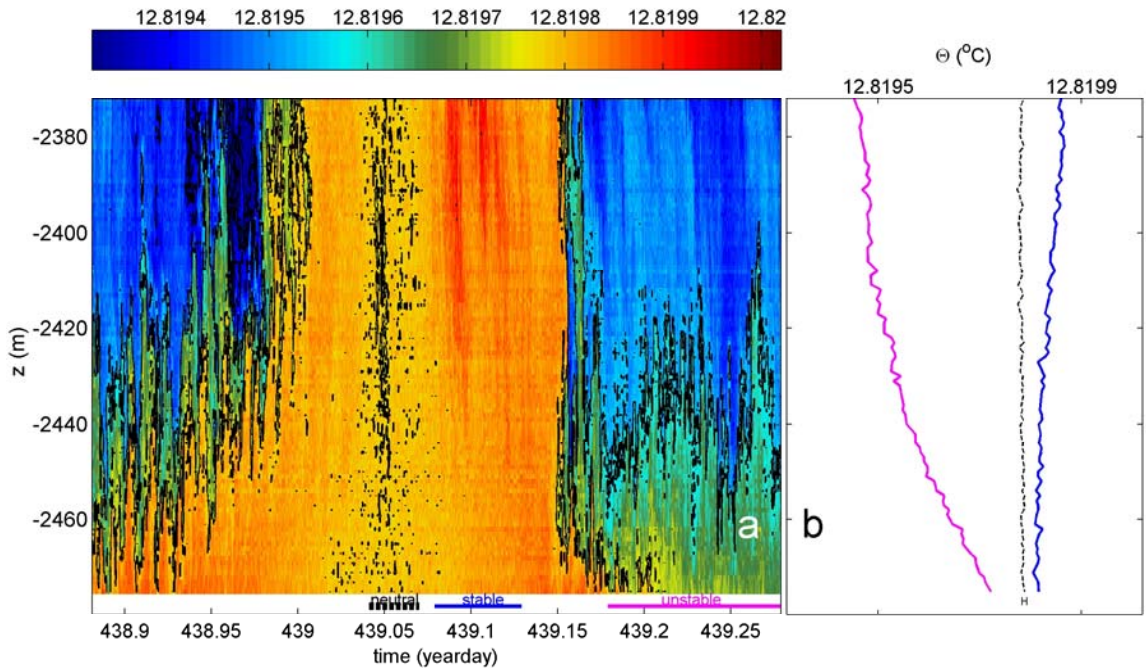

**Fig. 5** A >105-m tall near-homogeneous water column of which the mean-value profile from the dashed-black-bar-indicated period is used for reference of T-sensor data from other periods. (a) Time-vertical plot of Conservative Temperature from 1.0-m sampled line-c, lpf at 1000 cpd. The seafloor is at the level of the horizontal axis. Black contours are drawn every 0.0002 °C. The horizontal bars above the time axis indicate periods of which the mean values are given in b. They either reflect a period of neutral (homogeneous; black-dashed), statically stable (blue) or unstable (purple) conditions.

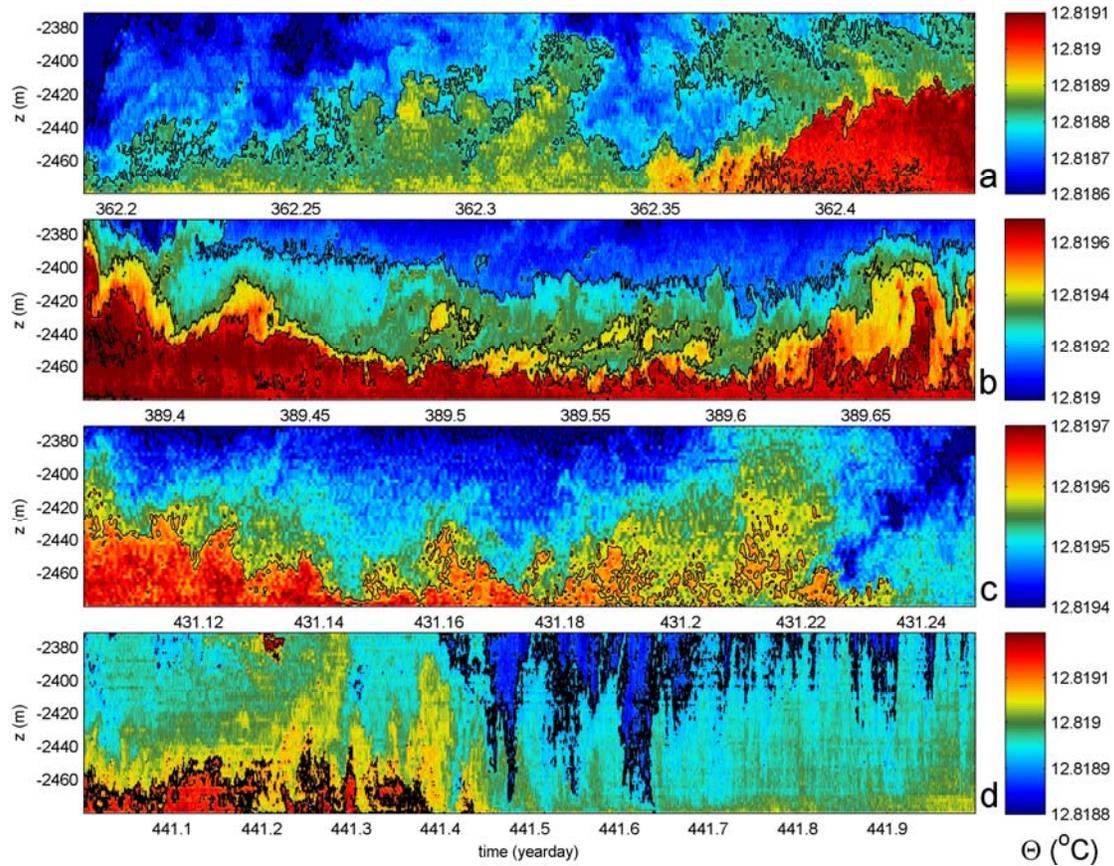

**Fig. 6** Magnifications' glossary of convection-turbulence of varying intensity and varying duration observed in the deep Western Mediterranean. The periods of the time-vertical plots of Conservative Temperature from 2.0-m sampled corner-line-1, lpf at 1000 cpd, are indicated by purple ticks in Fig. 4d. The seafloor is at the horizontal axes. Black contours are drawn every 0.0002 °C. (a) Early winter, total duration of 0.22 day, entire temperature variation of ΔΘ = 0.0005 °C. (d) Mid-winter, 0.38 day (half an inertial period), 0.0007 °C. (c) Late winter, 0.15 day, 0.0003 °C. (d) Late winter, 1.0 day, 0.0004 °C.

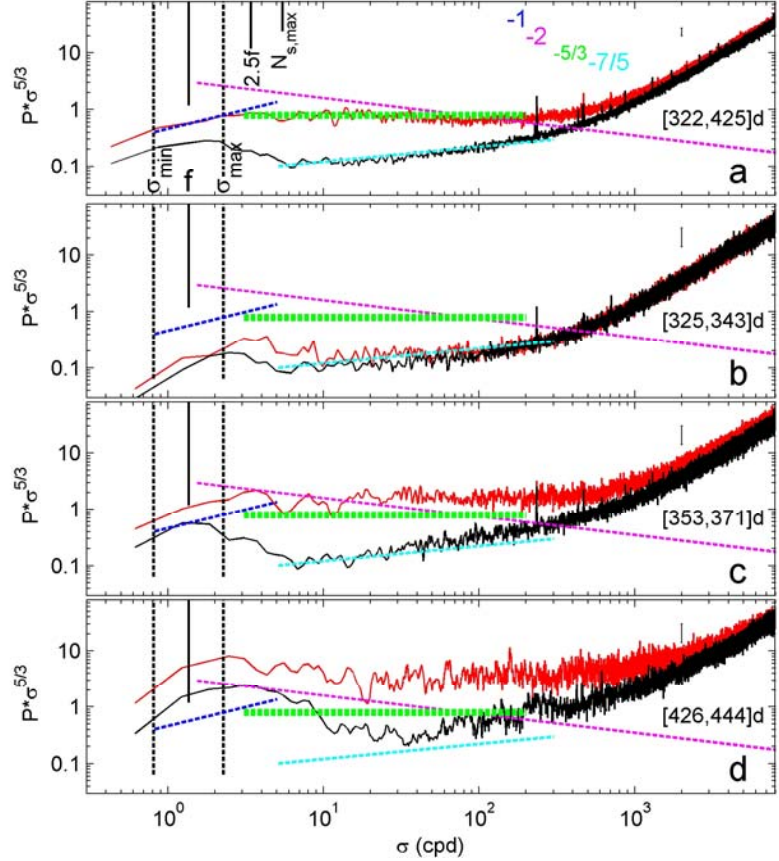

**Fig. 7** Log-log plots of moderately smoothed spectra for temperature variance scaled with (divided by) the inertial subrange slope with frequency ($\sigma$) of $\sigma^{-5/3}$. Averages are given for lower T-sensors from z = -2475 and -2459 m (black) and for upper T-sensors from -2393 and -2377 m (red). For reference, several spectral slopes are given with, e.g., '-1' indicating $\sigma^{-1}$ (in an unscaled plot). Short vertical black-solid lines indicate the inertial frequency f, 2.5f and average maximum small-scale buoyancy frequency $N_{s,max}$ = 4f. Long vertical dashed lines indicate the non-traditional inertio-gravity wave band [$\sigma_{min}$<f, $\sigma_{max}$>N] (LeBlond and Mysak 1978) under weakly stratified conditions N = f. (a) Averages for 103-day winter period before increased water-flow speeds due to sub-mesoscale eddy activity reach near the seafloor (cf., Fig. 4a). (b) Averages for 18-day early-winter period with low water-flow speeds of <0.1 m s$^{-1}$ and N ≈ 2f stratified waters near the seafloor. (c) As b., but one month later during a period with relatively large temperature instabilities. (d) As b., but for late-winter period with relatively strong water-flows and eddy activity.

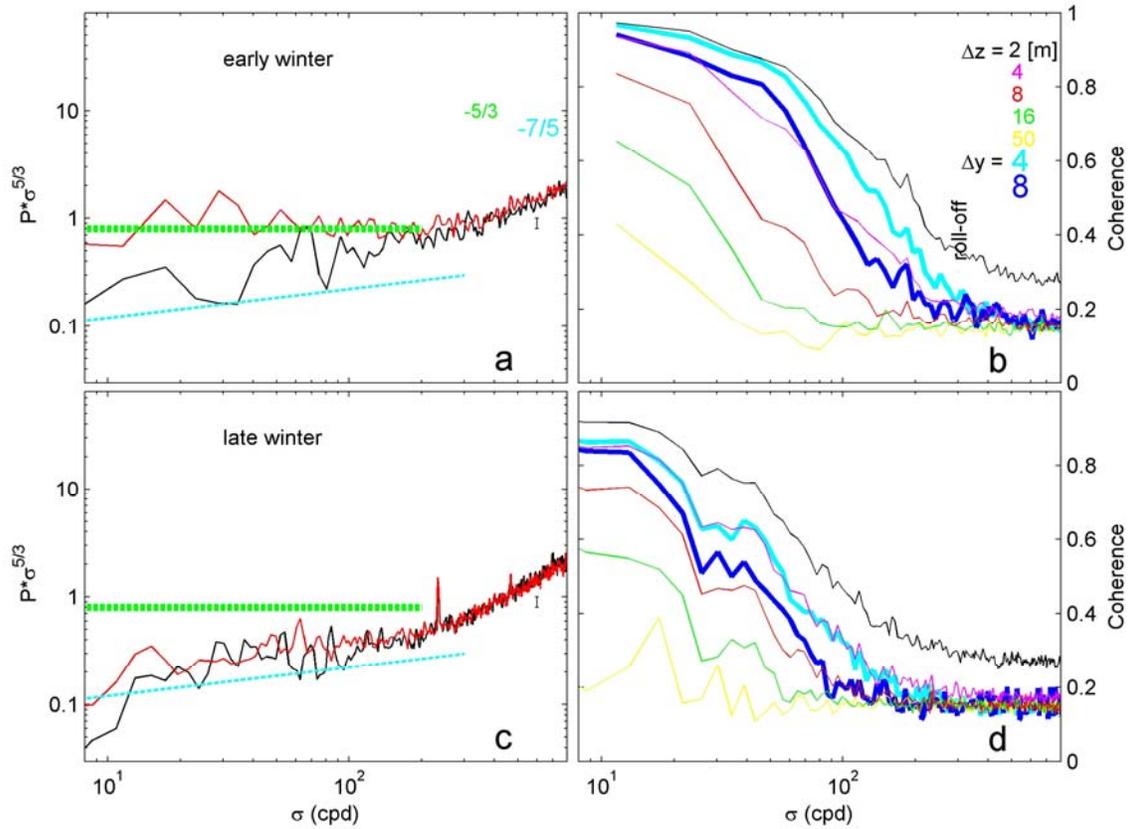

**Fig. 8** Spectral information on the transition from anisotropic to isotropic turbulence. Spectra are computed for two short periods of dominant geothermal heating: of 0.35 day in early-winter of Fig. 6a (a, b) (extended by 1 h on both sides for statistical reasons) and 1 day in late-winter of Fig. 6d (c, d). (a), (c) Similar to Fig. 7 but over a reduced x-axis between super-buoyancy and noise-roll-off frequencies. Temperature variance averages are computed for 50 upper T-sensors (10 per line) in red and for 50 lower T-sensors in black. (b), (d) Coherence spectra for all possible independent T-sensor pairs across indicated North-South horizontal Δy (thick blue lines) and vertical Δz separation distances. The 95% significance level is at about coh = 0.17.

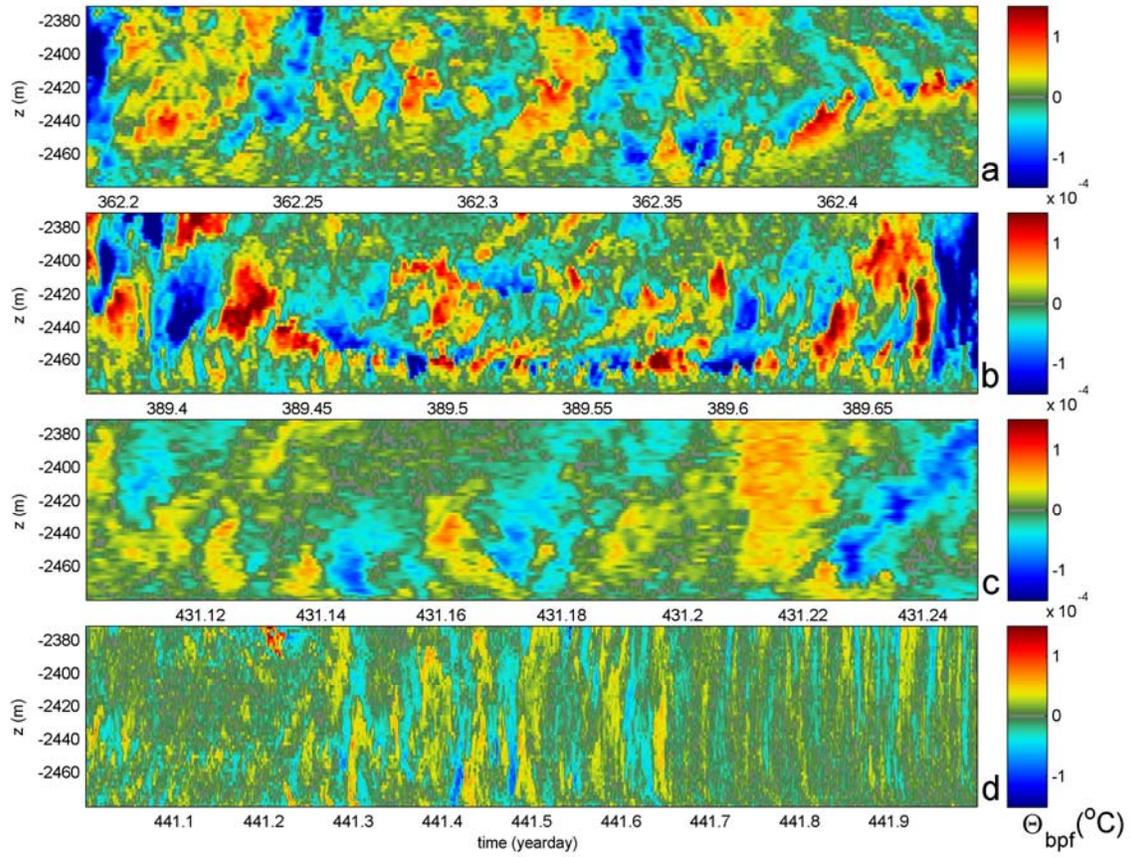

**Fig. 9** As Fig. 6, but for filtered data from passband [20, 200] cpd of turbulence range. Note that in this plot the colour ranges of temperature variation are identical for all panels.

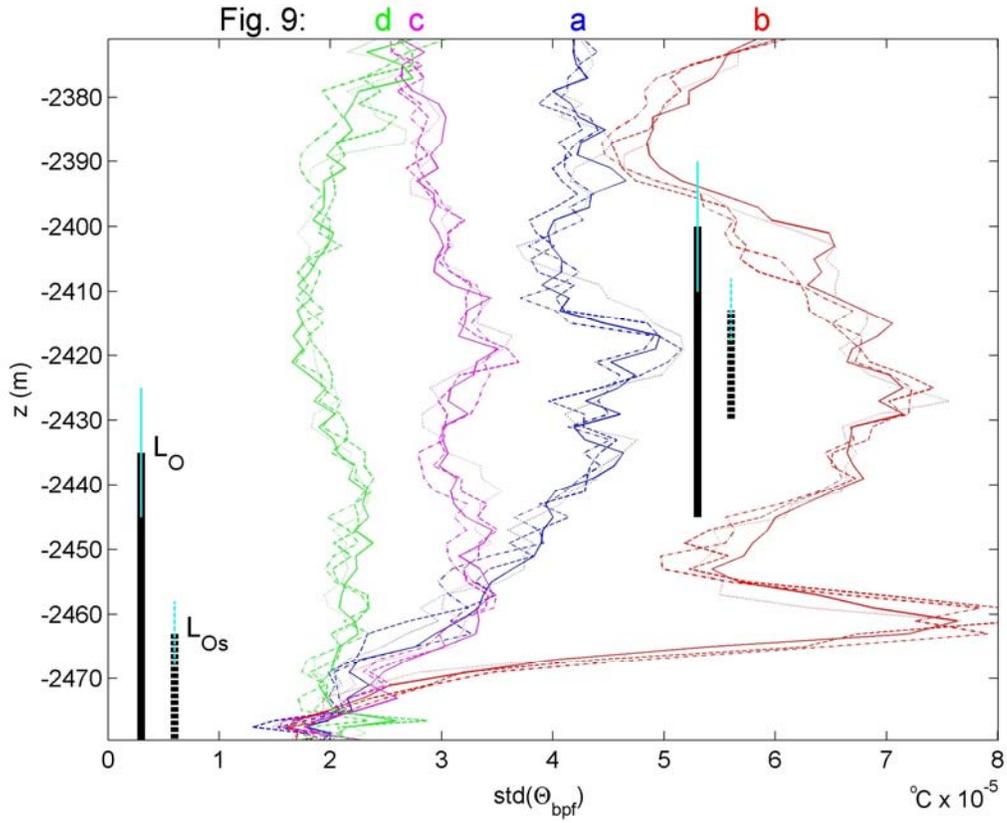

**Fig. 10** Vertical profiles of standard deviations of the data in Fig. 9, for corner-lines 1 (solid), 2 (dotted), 3 (dashed) and 4 (dash-dotted). Root-mean-square mean (large-scale) Ozmidov scale $L_O$ and small-scale $L_{Os}$ (explained in Section 4) are indicated by vertical bars in black solid and dashed, respectively, with corresponding error bars in thin light blue.